\documentclass[trackchanges,twocolumn]{aastex7}
\usepackage{amsmath}
\usepackage{rotating}
\usepackage{afterpage}
\usepackage{soul}
\usepackage{svg}
\usepackage{float}
\begin{document}

\title{The Influence of Clouds and Deuterium-Burning on Brown Dwarf Habitable Zones}

\author[orcid=0009-0001-9981-0171, sname='Smith']{Kayla J. Smith}
\altaffiliation{The University of Arizona}
\affiliation{Department of Planetary Sciences, Lunar and Planetary Laboratory, University of Arizona, Tucson, Arizona}
\email[show]{kaylasmith1@arizona.edu}  

\author[orcid=0000-0002-5251-2943, sname='Marley']{Mark S. Marley} 
\altaffiliation{The University of Arizona}
\affiliation{Department of Planetary Sciences, Lunar and Planetary Laboratory, University of Arizona, Tucson, Arizona}
\email{marksmarley@arizona.edu}

\begin{abstract}

To better understand the potential habitability of planets orbiting brown dwarfs, this work presents a new set of equilibrium temperature evolution tracks. Unlike most previous work that relied on analytic scaling relationships for brown dwarf luminosity evolution, we use the outputs of modern brown dwarf evolution models that account for the effects of deuterium burning, cloud formation and dissipation, and the most recent atmospheric opacities. While clouds are present, brown dwarfs cool more slowly than if they did not have clouds, allowing orbiting planets to remain in the habitable zone (HZ) for millions of years longer than previously estimated. Similarly, we find that during the deuterium-burning phase of brown dwarfs, which also slows the evolution, planets at the same orbital radius but orbiting brown dwarfs of different masses can remain in the HZ for the same duration, creating deuterium ``sweet spots'' for habitability around brown dwarfs near the deuterium-burning limit. For example, at 0.01 au a planet orbiting both a 0.012 and a 0.020 {M}$_\mathrm{\odot}$ brown dwarf stays in the HZ for $\sim 170 - 180\,\rm Myr$ because deuterium-burning more strongly affects the cooling of lower-mass brown dwarfs. The size of the effect decreases with decreasing orbital radius, with larger orbital radii having a more pronounced deuterium burning influence. These effects are absent from the analytic cooling approximations used in prior studies of substellar HZs and are revealed by our application of modern substellar evolution models.

\end{abstract}
\keywords{\uat{Habitable zone}{696} --- \uat{Brown dwarfs}{185}}


\section{Introduction} 
\label{sec:1}
\indent Most studies of planetary habitability consider planets orbiting main-sequence stars, like Earth orbiting the Sun. However, planets also form and orbit around brown dwarfs \citep[e.g.,][]{Zhou_2016, Fontanive_2020, Grieves_2021}. Brown dwarfs are substellar objects with masses below the hydrogen-burning limit ($\sim 0.073\,\rm M_\odot$)  but above the deuterium-burning limit ($\sim0.012\,\rm M_\odot$) \citep[][]{morley2024sonorasubstellaratmospheremodels}. Brown dwarfs occupy the mass range between giant planets and low-mass stars, with sufficient mass to briefly burn deuterium in their cores and effective temperatures spanning $\sim3000$ K and 250 K \citep[e.g.,][]{Burrows97, Spiegel_2011, Helling_2014}. Unlike main-sequence stars, which are massive enough to fuse hydrogen in their cores, brown dwarfs get cooler and less luminous as they age.

Brown dwarfs are nearly as abundant as main-sequence stars in the Universe, making the study of planets around these substellar objects crucial for expanding our understanding of habitability beyond our solar system. This is especially significant considering that M dwarfs constitute the lowest-mass and most abundant class of main-sequence stars in the Universe \citep{Kaminski_2025}. Observations have revealed that brown dwarf disks exhibit all three early stage planet formation processes: dust grain growth, crystallization, and settling, indicating that even these very low-mass objects possess the fundamental building blocks for planet formation \citep{2005Sci...310..834A}, and disks of stellar and substellar mass may be in a fragmentation-limited growth regime in which grain evolution depends on the host star's mass \citep{2016ApJ...831..125P}. Theoretical models predict that brown dwarfs with masses around 0.05 M$_\odot$ could form planets up to $\sim$5 M$_\oplus$ at orbital distances of $\sim$1 au, though the frequency of planet formation depends critically on disk mass. Earth-mass planets may be relatively common if brown dwarf disks are several Jupiter masses but extremely rare if typical disks are only a fraction of a Jupiter mass \citep{2007MNRAS.381.1597P}.

The characteristic that brown dwarfs cool over time, paired with their unique molecular opacity-dominated spectra affects the potential habitability of any planets that orbit them. The thermal emission spectra of brown dwarfs are punctuated by strong molecular features due to water, methane, and other molecules. Thus the atmospheric structure of any habitable planet with an atmosphere will likewise be shaped by the interaction with and absorption of this incident flux. Here, for simplicity, we neglect these spectral features and instead focus on the equilibrium temperature---arising from the integrated incident flux---of planets orbiting brown dwarfs to understand their habitable zone (HZ) lifetimes.

A main-sequence star can maintain a planet in the HZ for billions of years, although the HZ moves outward as the star ages \citep[e.g.,][]{Kopparapu_2013}. This is because such stars get warmer and more luminous over time. On the other hand, a brown dwarf's HZ moves inward over time as the brown dwarf cools. The cooling proceeds nonuniformly with time, influenced by episodic processes including deuterium burning and the formation or dissipation of condensate clouds, each of which alters the rate of energy loss. 

The deuterium-burning threshold in brown dwarfs falls approximately in the range of 0.0114 and 0.0144 {M}$_\mathrm{\odot}$ \citep[e.g.,][]{Spiegel_2011, Marley_2021, morley2024sonorasubstellaratmospheremodels} with the specific limit depending on the details of the atmospheric opacity. A value of 0.013 {M}$_\mathrm{\odot}$ is  frequently relied upon as the standard. In the models employed in this study, deuterium burning occurs in brown dwarfs with masses greater than approximately 0.012 to 0.0125 {M}$_\mathrm{\odot}$, for both cloudy and cloudless evolutionary tracks. Brown dwarfs exceeding 0.020 {M}$_\mathrm{\odot}$ will burn through all their deuterium quickly, within 20 million years of their formation, whereas objects below 0.011 {M}$_\mathrm{\odot}$ lack sufficient mass to sustain any appreciable deuterium fusion \citep[e.g.,][]{Morley_2019, Saumon_1996, Spiegel_2011}. \citep{morley2024sonorasubstellaratmospheremodels}. Deuterium burning in the mass interval between about 0.012 and 0.020 {M}$_\mathrm{\odot}$ is interesting as it happens more slowly and occurs  later in the evolution of brown dwarfs than in more massive objects, making for intriguing effects on HZ lifetimes. 

Furthermore, clouds form in brown dwarf atmospheres as they cool  
\citep[e.g.,][]{Helling_2014,Marley_2015}. The atmospheric temperature profiles of brown dwarfs span the condensation points of multiple compounds, resulting in their clouds of various compositions forming and sinking over time  as the objects age \citep[e.g.,][]{Marley02, https://doi.org/10.1002/asna.201211776}. Such clouds alter the atmospheric thermal structure and cooling rates of brown dwarfs as well as their emergent spectra. Optically thick clouds cause the brown dwarf to cool off slower over time than if clouds were not present \citep[e.g.,][]{Cooper_2003}. Several models have been developed to study cloud structure in brown dwarf atmospheres \citep[e.g.,][]{Ackerman_2001}, providing insights into how clouds influence their thermal structure, spectral properties, and overall evolution \citep[e.g.,][]{MarleySaumon2008,Lef_vre_2022,morley2024sonorasubstellaratmospheremodels}. These studies all document how thermal evolution slows during the time that cloud opacity is most abundant in brown dwarf atmospheres.

\begin{figure*}[!htb]  
    \centering
    \includegraphics[scale=0.5]{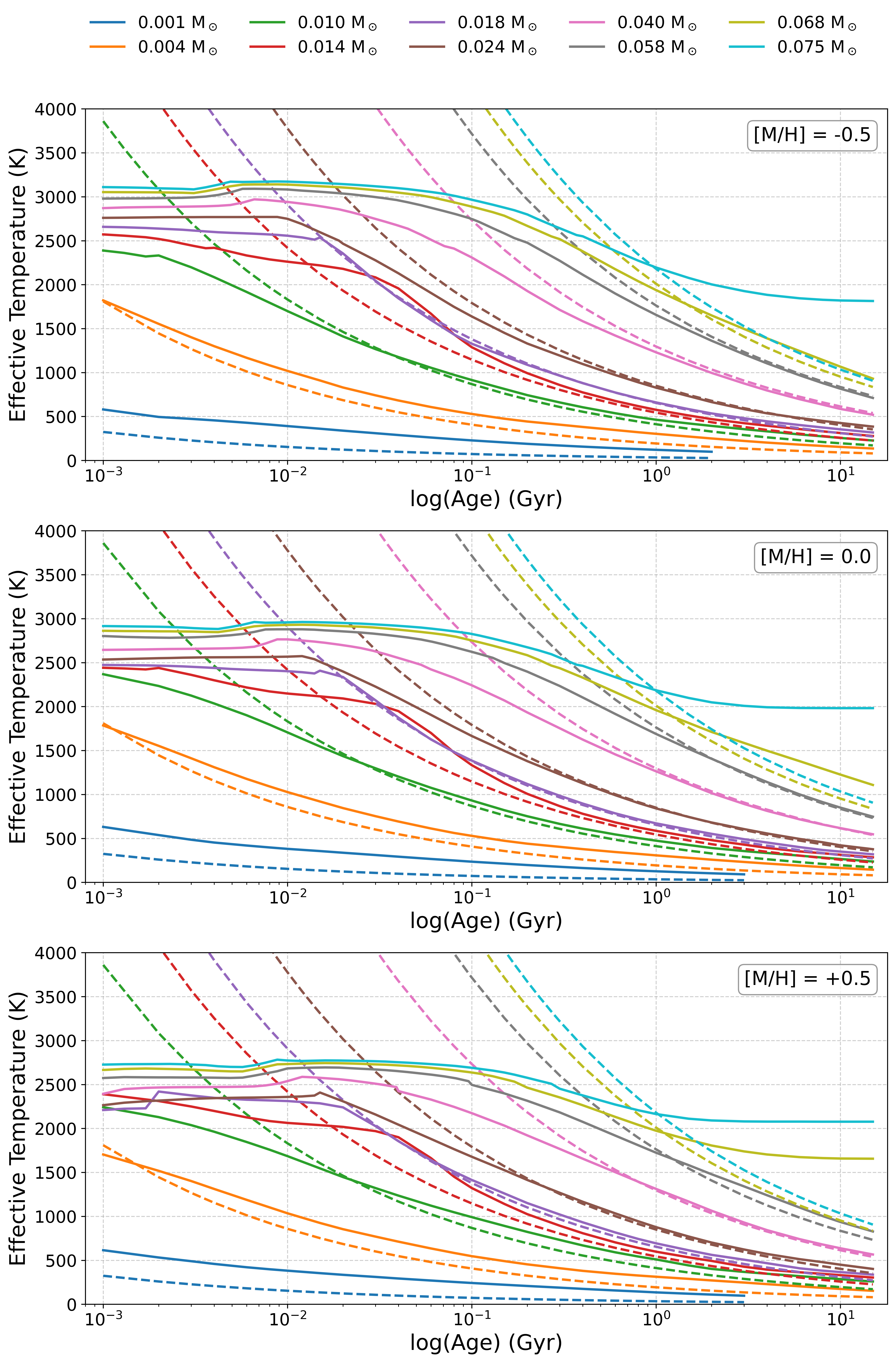}
    \caption{Brown dwarf evolution for different masses (lines) and metallicities (panels). The solid lines show the cloudless evolutionary models used in this work, which combine the Red Diamondback and Bobcat models \citep[][]{Marley_2021, 2025arXiv251008694D}, as described in Section 2. The dashed lines are the analytical predictions using Equation~\ref{eq:1} by \cite[][]{Lingam_2020}. The top, middle, and bottom panels are illustrating brown dwarfs of metallicities -0.5, +0.0, and +0.5 respectively.}
    \label{fig:loeb}
\end{figure*}

The nature and extent of brown dwarf HZ have been studied \citep[e.g.,][]{Lingam_2020, Bolmont_2018, Barnes_2013, 2004IAUS..213..115A} but most of this body of work has only utilized power-law fits to brown dwarf evolution model predictions rather than the models themselves to compute equilibrium temperatures for planets orbiting brown dwarfs. For example, \cite{Lingam_2020} used the following expression to approximate the effective temperature of brown dwarfs as a function of time:
\begin{equation}
T_{\text{BD}} = 59 \, \text{K} \times \left(\frac{t_{\text{BD}}}{1\, \text{Gyr}}\right)^{-0.324} \times \left(\frac{M_{\text{BD}}}{1\,{M_J}}\right)^{0.827}
\label{eq:1}
\end{equation}
where \textit{t}$_\mathrm{BD}$ is the age of the brown dwarf, \textit{M}$_\mathrm{BD}$ is the mass of the brown dwarf, and \textit{M}$_\mathrm{J}$ is the mass of Jupiter. This approximation is from the brown dwarf evolution model of \cite{burrows1993science}, developed before the first undisputed discovery of a brown dwarf. The power-law fit does not capture structure from deuterium burning or cloud formation and dissipation in the evolutionary behavior and produces unrealistically high temperatures at young ages. Differences between this power-law model and the models used here are shown in Figure~\ref{fig:loeb}.   

Similarly, \cite{Bolmont_2018} studied the tidal evolution of a planet orbiting a brown dwarf, but their brown dwarf cooling models did not fully account for the effects of cloud dynamics in the luminosity evolution of the parent object. \cite{Barnes_2013} used models of \cite{2003A&A...402..701B} for the evolution of brown dwarf luminosity over time to understand habitability.

Here, we aim to refine our understanding of HZs around brown dwarfs, and compare these conclusions with the conclusions reached in the above studies. Section~\ref{sec:2} details the key equations and outputs necessary for the analysis. In Section~\ref{sec:3}, we present our findings across four areas: the dependence of planetary equilibrium temperature on orbital distance, the impact on HZ lifetime as a function of brown dwarf mass and deuterium-burning, the influence of metallicity on planetary HZ lifetimes, and the overall HZ limit evolution over time. The discussion and conclusions are in sections ~\ref{sec:4} and ~\ref{sec:5}, respectively.

\section{Methods} 
\label{sec:2}
We use a variety of models for the evolution of brown dwarf luminosity through time. The most important parameter affecting substellar evolution is clouds as they add continuum atmospheric opacity, impeding thermal emission of energy from the atmosphere and slowing cooling times. Because clouds are difficult to model, we consider two different cases for the HZ calculation to understand the impact of brown dwarf clouds on planet habitability. In the limit where clouds are never present, we employ the cloudless Sonora “Bobcat” evolution models \citep{Marley_2021}. These cloudless cooling tracks provide a lower limit to HZ lifetimes. These are also most similar at later ages to the models used in previous work, providing a point of comparison. 

In reality, all brown dwarfs have clouds at some point in their lifetime, forming predominantly below about 2200 K. The Sonora ``Diamondback” grid is used for the cloudy brown dwarf evolution \citep{morley2024sonorasubstellaratmospheremodels}. The brown dwarfs in this model set initially include cloud opacity which slows the cooling relative to the Bobcat evolution, but at $\sim1300$ K, the clouds are assumed to dissipate, further affecting the cooling. While a number of cloud treatments are possible, the difference between the cloudless tracks and the adopted Diamondback evolution tracks will be much greater than differences that would arise from different treatments of the cloud opacity with cloudy evolution (e.g., various cloud model choices).

Because both the Diamondback and Bobcat models do not capture very early brown dwarf evolution ($T_{\rm eff} > 2400\,\rm K$) we use the ``Red Diamondback" evolution models \citep{2025arXiv251008694D} that rely on the SPHINX stellar evolution models \citep[][]{Iyer_2023} for this period of cooling, allowing us to begin evolving all masses at the same age of 1 Myr. At $T_{\rm eff} = 2400\,\rm K$  we smoothly transition from Red Diamondback to either the Bobcat or Diamondback tracks, as the case may be. Since clouds do not impact the evolution above 2400 K, the transition to either the Bobcat or Diamondback models is smooth. Overall, using both of those model sets allows for a direct comparison between cloudless and cloudy brown dwarf evolution, with the latter offering a more realistic representation of their true behavior.

Figure~\ref{fig:loeb} compares the combined Bobcat and Red Diamondback evolution models at three metallicities to the evolution models used in \cite{Lingam_2020}. Note that while the evolution is similar at older ages (greater than 1 Gyr), at young ages the Lingam et al. fitting formulae give effective temperatures that are much hotter than the Red Diamondback evolution. This is especially relevant to planets in small orbits that, with the new models, never receive exceptionally high incident fluxes.

The panels in Figure~\ref{fig:cloudy_temp} compare the combined cloudless evolution with the cloudy evolution. The effect of the slower cloudy evolution is apparent at later ages as it takes longer for the cloudy models to reach the same effective temperature. This means that a given equilibrium temperature is reached later for a planet orbiting brown dwarfs following the cloudy cooling tracks.

We follow the usual approach to computing the equilibrium temperature for a planet orbiting a star: 

\begin{equation}
T_{\text p} = T_{\text{BD}} \sqrt{\frac{R_{\text{BD}}}{2a}} (1 - A_{\text p})^{1/4}
\label{eq:2}
\end{equation}
\noindent where $T_{\text{BD}}$ is the effective temperature of the brown dwarf, $R_{\text{BD}}$ is the radius of the brown dwarf, $a$ is the orbital radius, and $A_{\text{p}}$ is the planet Bond albedo. Throughout this work, `habitability' refers strictly to the radiative HZ defined by equilibrium temperature and does not account for atmospheric, geophysical, or tidal processes that may strongly modify surface conditions. The temperature range we consider is 175-275 K, similar to \citep[e.g.,][]{Kopparapu_2013}. Here we assume a constant Bond albedo of 0.25, but in reality, $A_{\rm p}$ will vary as the brown dwarf spectrum changes in relation to the planet's atmospheric absorption profile with time \citep[][]{1999ApJ...513..879M}. A comparison of different Bond albedos is presented in the following section. We accept this approximation for now as we are instead focused on the effect of the brown dwarf luminosity evolution. In the future we will consider the impact of the changing shape of the emergent spectra with time. 

Using each evolution model set, we computed tables of $T_{\rm p}$ of planets orbiting brown dwarfs of various masses and various orbital radii and then  interpolated in these tables to determine the age at which each planet reached 275 K (inner HZ) and 175 K (outer HZ). By subtracting these two ages, the lifetime (or the time a planet resides in the HZ) of each planet was calculated.

\begin{figure*}[!htb]  
    \centering
    \includegraphics[scale=0.6]{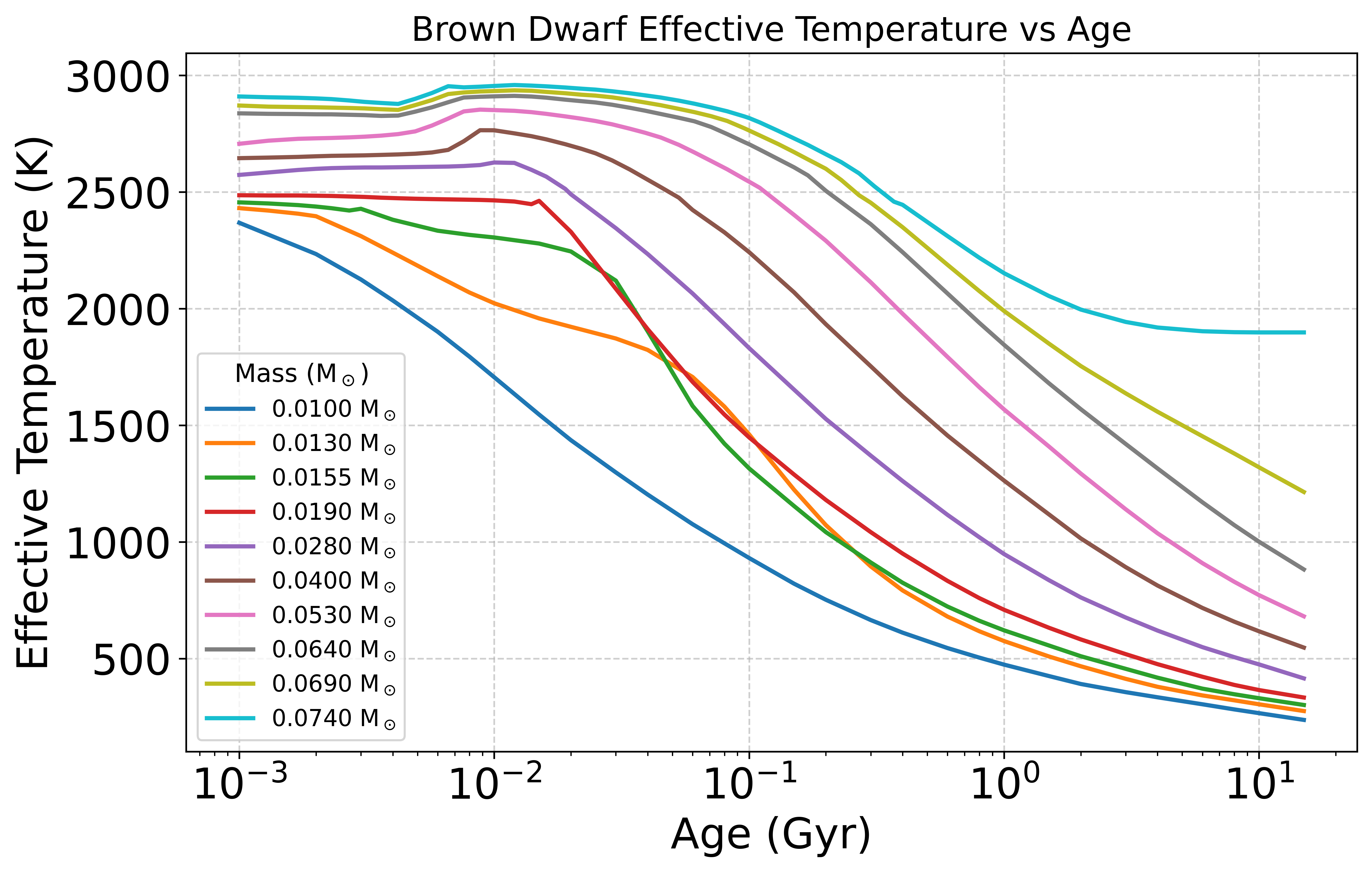}
    \label{fig:cloudless_temp}
\end{figure*}

\begin{figure*}[!htb]  
    \centering
    \includegraphics[scale=0.6]{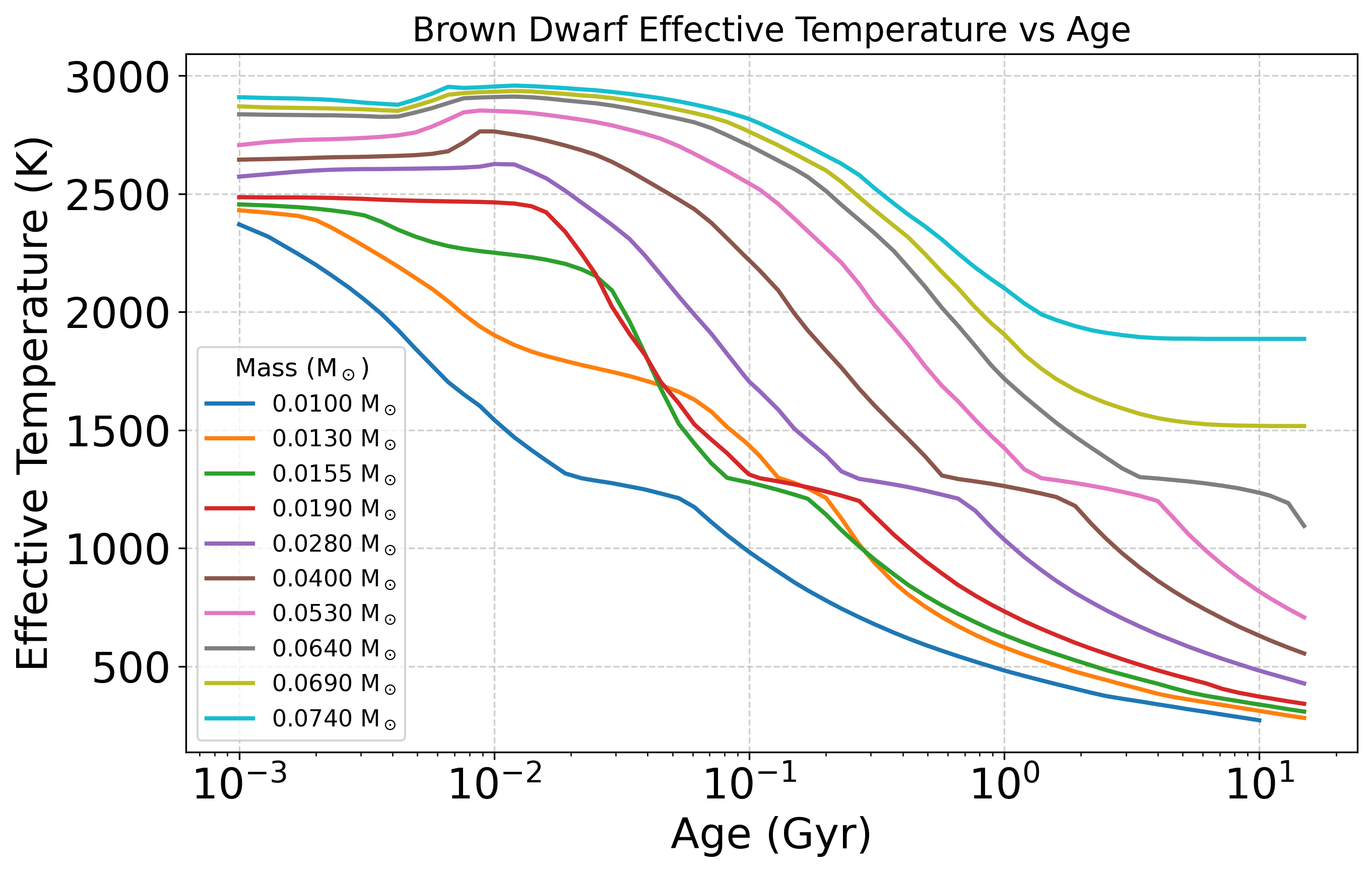}
    \caption{Evolution of brown dwarf effective temperature as in Figure~\ref{fig:loeb}, but comparing cloudless evolution (top panel; \citep[][]{Marley_2021} and cloudy evolution (bottom panel; \citep[][]{morley2024sonorasubstellaratmospheremodels} evolution as described in the text. Models in both panels have [M/H] = $+0.0$.}
    \label{fig:cloudy_temp}
\end{figure*}

\section{Results}
As stated in section~\ref{sec:1}, the following section provides an overview of the results across four regimes: equilibrium temperature, HZ lifetime as influenced by brown dwarf deuterium burning, HZ lifetime as a function of host brown dwarf metallicity, and the evolution of the brown dwarf HZ size.
\label{sec:3}
\subsection{Equilibrium Temperature}
\label{sec:3.1}
\begin{figure*}[!htb]  
    \centering
    \includegraphics[scale=0.7]{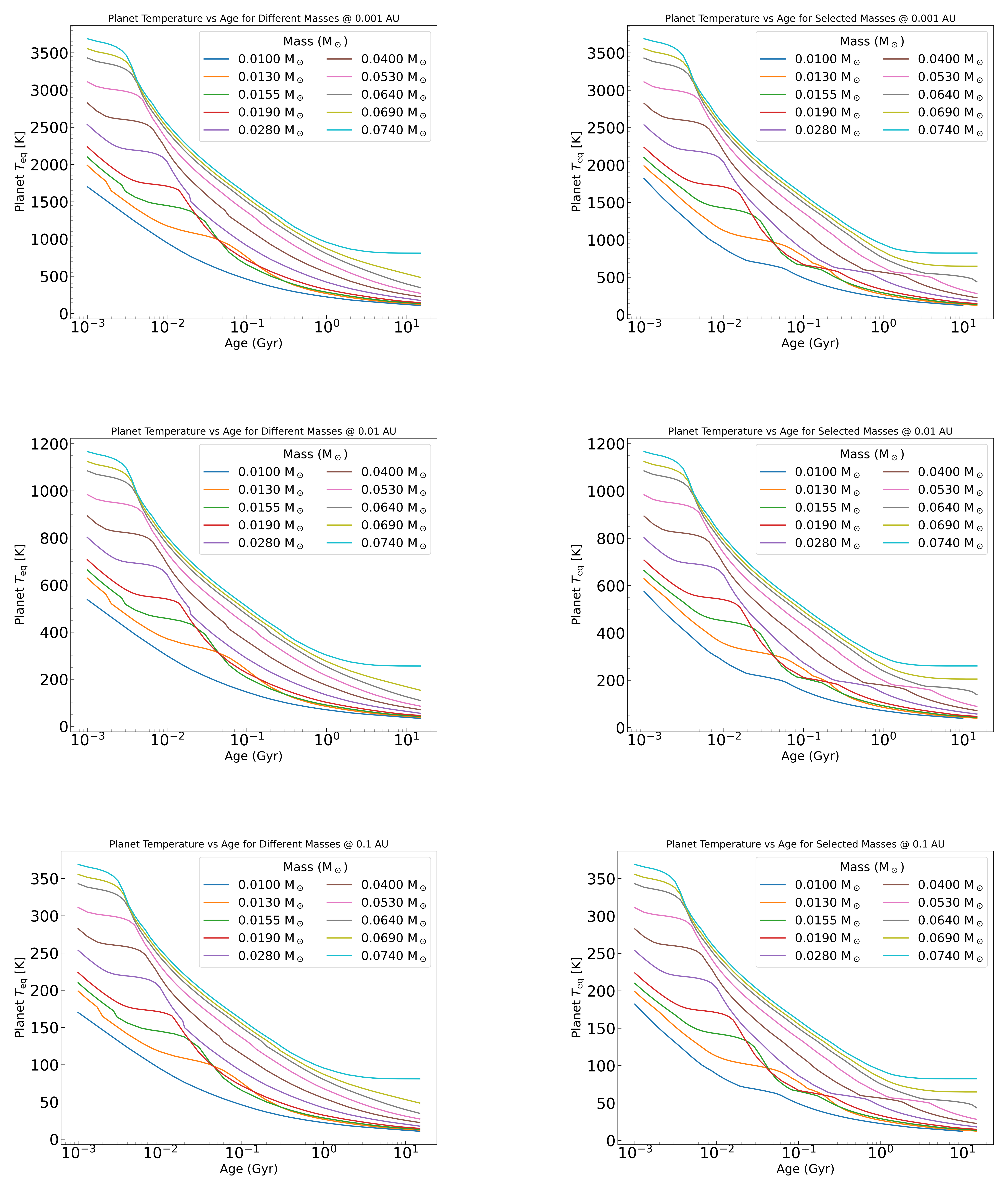}
    \caption{Planet equilibrium temperature versus log age of the host brown dwarf for different orbital radii. Panels in the left column present $T_{\rm eq}$ assuming the cloudless evolution while the right column is for the cloudy evolution. All of the planet temperatures were calculated orbiting a brown dwarf with a [M/H] = $+0.0$ and Bond albedo of 0.25.}
    \label{fig:equilibrium}
\end{figure*}

To explore the dependence of equilibrium temperature on orbital radius with the adopted evolution models, we considered three orbital radii: 0.001, 0.01, and 0.1 au. Figure~\ref{fig:equilibrium} presents the computed planet equilibrium temperatures as a function of time for the different orbital radii and different brown dwarf evolution model cases. Figure~\ref{fig:equilibrium} shows that only brown dwarfs below $0.06,\rm M_\odot$ (cloudless evolution) and $0.05,\rm M_\odot$ (cloudy evolution) cool sufficiently for planets at 0.001 au to enter the HZ at ages $\gtrsim 1$ Gyr. This establishes a clear mass threshold for HZ access at very small orbital radii: more massive brown dwarfs remain too luminous for such planets to ever reach equilibrium temperatures below 275 K within a Hubble time.

Some planets begin their evolution well above the adopted upper HZ temperature limit (275 K). \cite{Lingam_2020} argued that such initially hot planets would likely lose their water before cooling into the HZ. However, their analytic power-law prescription for brown dwarf luminosity overestimates effective temperatures at young ages compared to modern evolutionary models. In our adopted evolution models, brown dwarfs are substantially cooler at early times, and for masses below $\sim0.053,\rm M_\odot$, planets orbiting at 0.01 au and beyond never exceed 1000 K. This significantly moderates the severity of early runaway heating relative to previous estimates. A full treatment of runaway is defined for future work.

Cloud formation further modifies the timing of HZ entry and exit. Because clouds slow brown dwarf cooling, planets orbiting cloudy evolution brown dwarfs remain warmer for longer and therefore enter and exit the HZ at later times than in the cloudless case. Importantly, clouds do not determine whether a planet can ever reach the HZ as that outcome is set primarily by brown dwarf mass and orbital radius. Instead, clouds regulate the timing and duration of habitable conditions once HZ access is possible.

\subsection{HZ Lifetime versus Brown Dwarf Mass and Studying Deuterium Burning}
\label{sec:3.2}
In this section, we examine the HZ lifetimes of planets orbiting both cloudless and cloudy brown dwarf evolution tracks, considering a range of brown dwarf masses at a range of orbital distances.  We identify deuterium-burning ``sweet spots'' for potential habitability of planets orbiting brown dwarfs. Notably, planets at the same orbital radius around brown dwarfs of different masses near the deuterium-burning limit can remain in the HZ for comparable durations. This phenomenon illustrates the significance of deuterium fusion, which temporarily enhances brown dwarf luminosity and extends the window for planetary habitability. We find that brown dwarfs with masses between 0.012 and 0.020 $M_\odot$ exhibit comparable HZ lifetimes at a fixed orbital distance. This similarity arises because deuterium burning prolongs the cooling timescale of the lower-mass object, an effect that we discuss in greater detail later in this section.

\begin{figure*}[!htb]  
    \centering
    \includegraphics[scale=0.85]{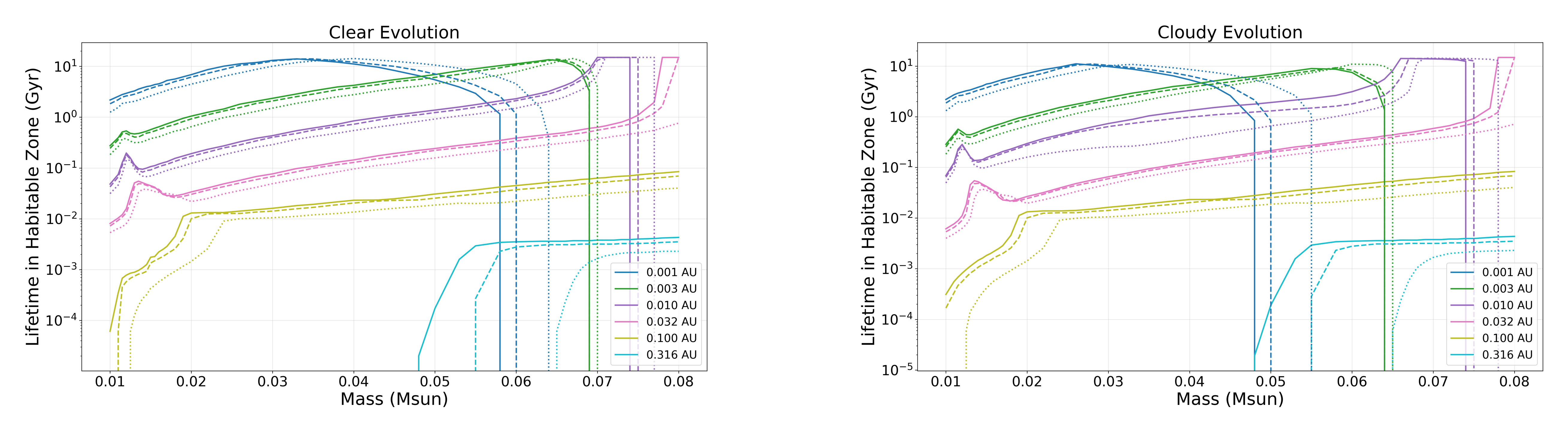}
    \caption{Planet equilibrium temperature versus log age of the host brown dwarf for difference orbital radii. The left panel is the amount of time planets orbiting cloudless brown dwarfs are $175 \le T_{\rm eq} \le 275\,\rm K$ K, or the HZ. The right panel is the same plot but with cloudy brown dwarf evolution. Each line corresponds to a log step orbital radius from 0.001 to 0.01. The solid, dashed, and dotted line correspond to a Bond albedo of 0.125, 0.25, and 0.50, respectively. Both cases have [M/H] = $+0.0$.}
    \label{fig:lifetime_cloudless_cloudy}
\end{figure*}

As shown in Figure~\ref{fig:lifetime_cloudless_cloudy}, as the orbital radius increases, the range of brown dwarf masses where planets spend a finite amount of time inside the HZ equilibrium temperature range varies. Around low mass brown dwarfs there is a finite time spent in the HZ at 0.001 au up to $0.06 \,\rm M_\odot$ for the clear and $0.05\,\rm M_\odot$ for the cloudy evolution, respectively. More massive brown dwarfs never cool off enough for the planet at this orbital radius to enter the HZ temperature range in the age of the Universe. However, for the lower brown dwarf masses planets at this orbital radius can spend well in excess of 1 Gyr in the HZ. Varying the Bond albedo from 0.125 to 0.50 demonstrates that planetary albedo directly affects the duration a planet remains in the HZ. In the deuterium-burning mass regime, planets with higher albedo spend less time in the HZ compared to their lower-albedo counterparts. However, higher-albedo planets can maintain habitable conditions around more massive brown dwarfs, with this effect being modulated by orbital radius.

At larger ($\ge$ 0.1 au) orbital distances, only the most massive brown dwarfs are ever warm enough to maintain a HZ planet, although it would just be for a few million years or less. At 0.032 au, planets have a finite habitable-zone lifetime around brown dwarfs of every mass. Because cloudy brown dwarf evolution proceeds more slowly, resulting in a more gradual cooling, planets irradiated by the cloudy evolution tracks remain in the habitable zone longer than the cloudless evolution case for the same mass. 

The impact of deuterium-burning is clearly visible in both panels of Figure~\ref{fig:lifetime_cloudless_cloudy}, appearing around $\sim0.012 {M}_\mathrm{\odot}$.  Although this phase is relatively short-lived in brown dwarfs, it has noteworthy implications for planetary habitability for planets at the right orbital radius. By slowing the cooling of planets near the deuterium burning limit, the process extends the duration a planet remains within the HZ. For example, a planet orbiting a 0.012 $M_\odot$ brown dwarf (with [M/H] = $+0.0$) at 0.01 au remains in the HZ for approximately 182 Myr, while a similar planet orbiting a 0.02 $M_\odot$ brown dwarf at the same distance remains habitable for about 170 Myr. Likewise, in the cloudy evolution case, a planet at 0.01 au around a 0.012 $M_\odot$ brown dwarf is habitable for roughly 272 Myr, compared to 275 Myr for one orbiting a 0.02 $M_\odot$ brown dwarf under the same conditions. Note that in comparing the cloudless versus cloudy evolution we find that the cloudy evolution results in longer HZ lifetimes  of $\sim10$ million years. 

For the deuterium burning phase to notably impact the HZ lifetime of a planet, the HZ equilibrium temperature must be reached neither too early nor too late in the brown dwarf evolution such that the effect of the cooling is either yet to appear or has already dissipated. This is why there is little impact on HZ lifetime by deuterium burning at 0.1 au and beyond.

\subsection{The Effects of Metallicity on Brown Dwarf HZs}
\label{sec:3.3}
\begin{figure*}[!htb]  
    \centering
    \includegraphics[scale=0.85]{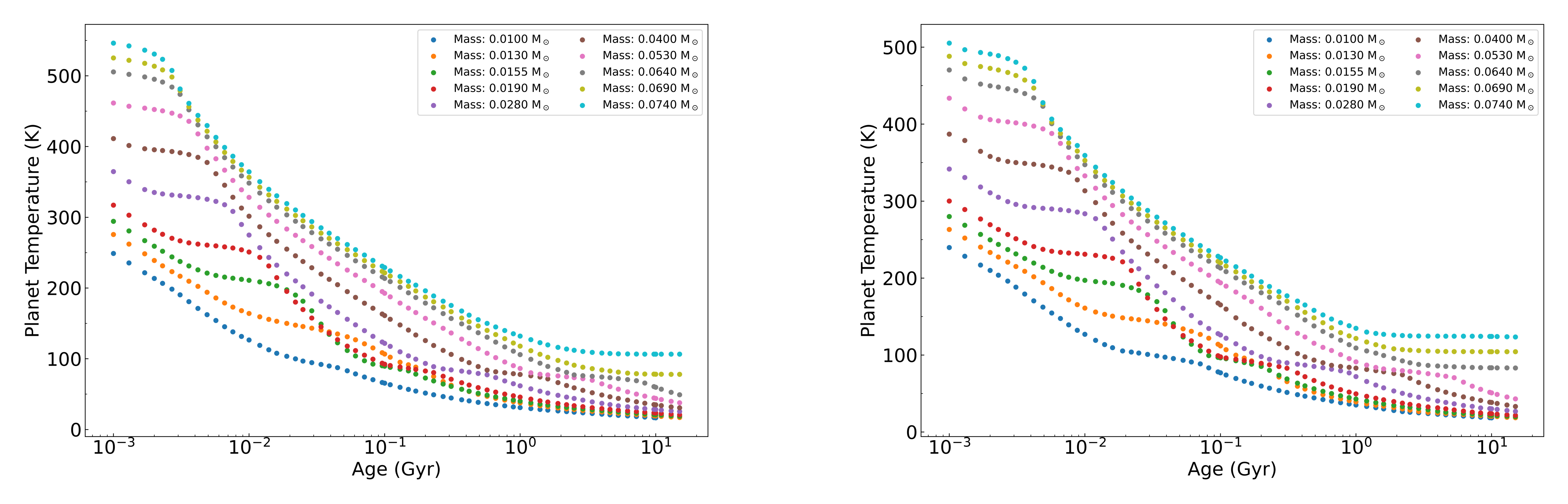}
    \caption{Planet equilibrium temperature (calculated using Equation~\ref{eq:2}) versus log age of the cloudy brown dwarf in Gyr. The left panel is [M/H] = $-0.5$ and the right panel is [M/H] = $+0.5$. The planets are orbiting at 0.05 au.}
    \label{fig:cloudy_planet_temp}
\end{figure*}

Our evolution models allow us to isolate the impact of brown dwarf metallicity on planetary equilibrium temperatures and HZ lifetimes. Figures ~\ref{fig:cloudy_planet_temp} illustrates the temperature–-age evolution for planets orbiting cloudy evolution brown dwarfs with [M/H] = $-0.5$ and $+0.5$ at 0.05 au. Higher-metallicity brown dwarfs cool more slowly and remain warmer at late times, while at very young ages they are slightly cooler than their lower-metallicity counterparts (also illustrated in Figure~\ref{fig:loeb}). The behavior at early times reflects differences during the collapse phase: lower metallicity reduces atmospheric opacity, allowing energy to escape more efficiently \citep[e.g.,][]{Omukai_2010}. At later times, however, the higher opacity of metal-rich atmospheres slows the long-term radiative cooling of the brown dwarf, allowing these objects to remain warmer than lower metallicity brown dwarfs, as they evolve.

\begin{figure*}[!htb]  
    \centering
    \includegraphics[scale=0.5]{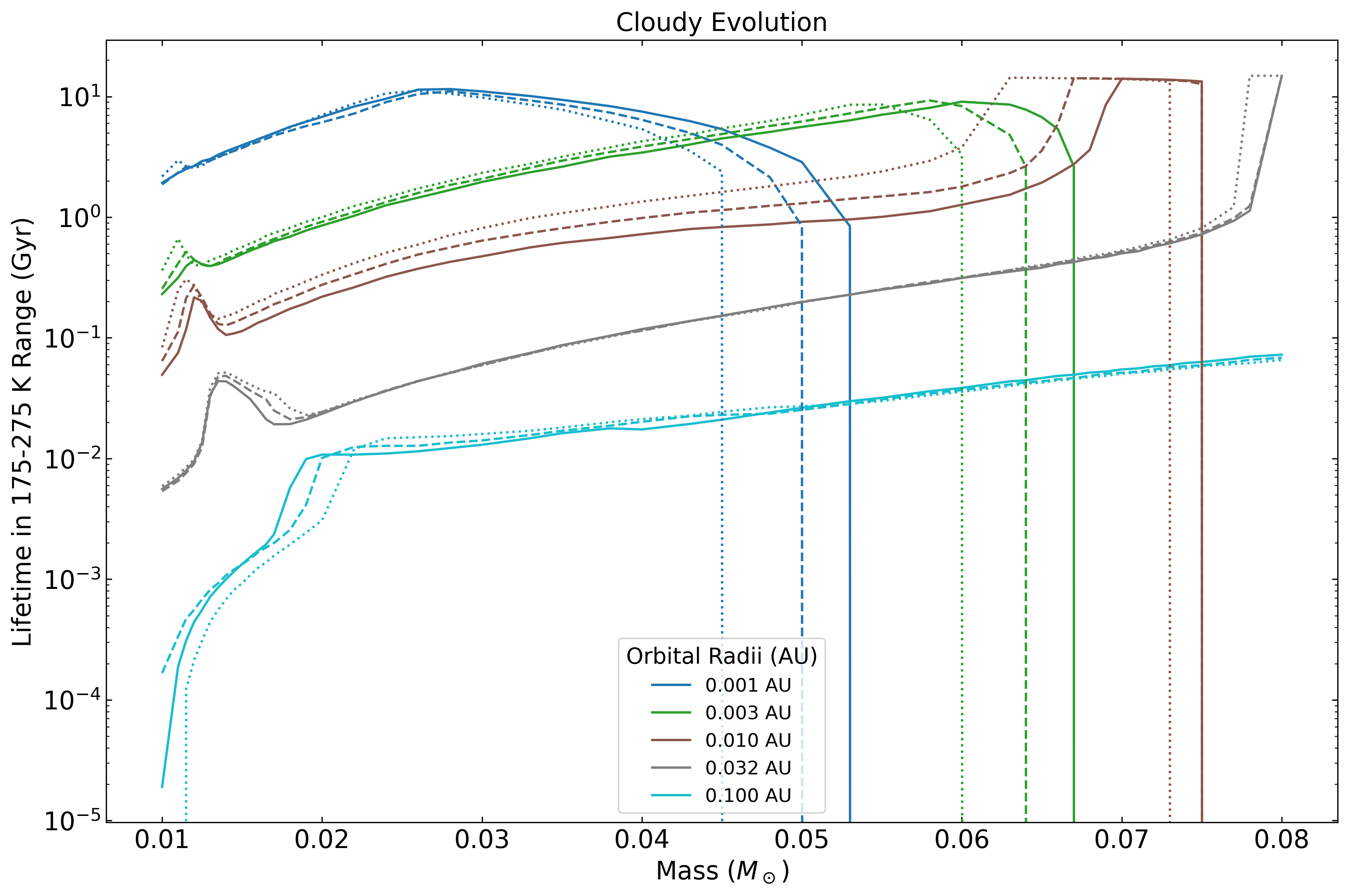}
    \caption{This figure shows the duration that planets orbiting cloudy brown dwarf evolution models remain in the HZ (y-axis) as a function of brown dwarf mass (x-axis). Each line represents a specific orbital radius, calculated using Equation~\ref{eq:2}. The solid line is a brown dwarf [M/H] = $-0.5$, the dashed line is a brown dwarf [M/H] = $+0.0$, and the dotted line is a brown dwarf [M/H] = $+0.5$.}
    \label{fig:lifetime_cloudy_fortney}
\end{figure*}

\begin{figure*}[!htb]  
    \centering
    \includegraphics[scale=0.85]{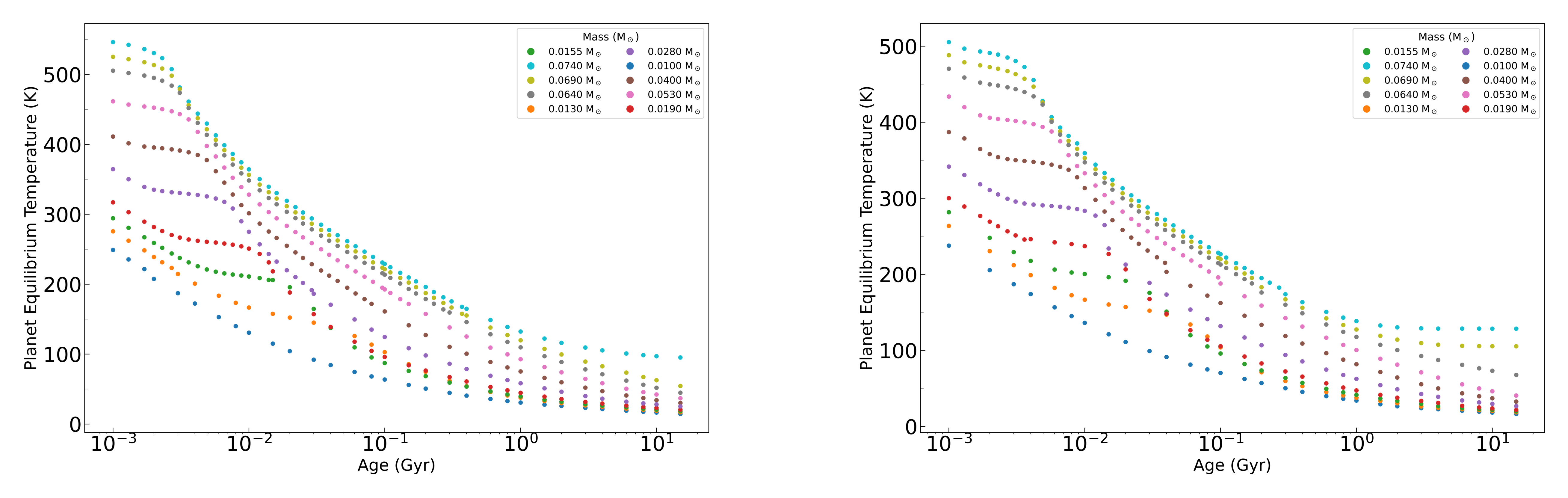}
    \caption{Planet equilibrium temperature (calculated using Equation~\ref{eq:2}) versus log age of the cloudless brown dwarf in Gyr. The left panel is [M/H] = $-0.5$ and the right panel is [M/H] = $+0.5$. The planets are orbiting at 0.05 au.}
    \label{fig:cloudless_planet_temp}
\end{figure*}

\begin{figure*}[!htb]  
    \centering
    \includegraphics[scale=0.5]{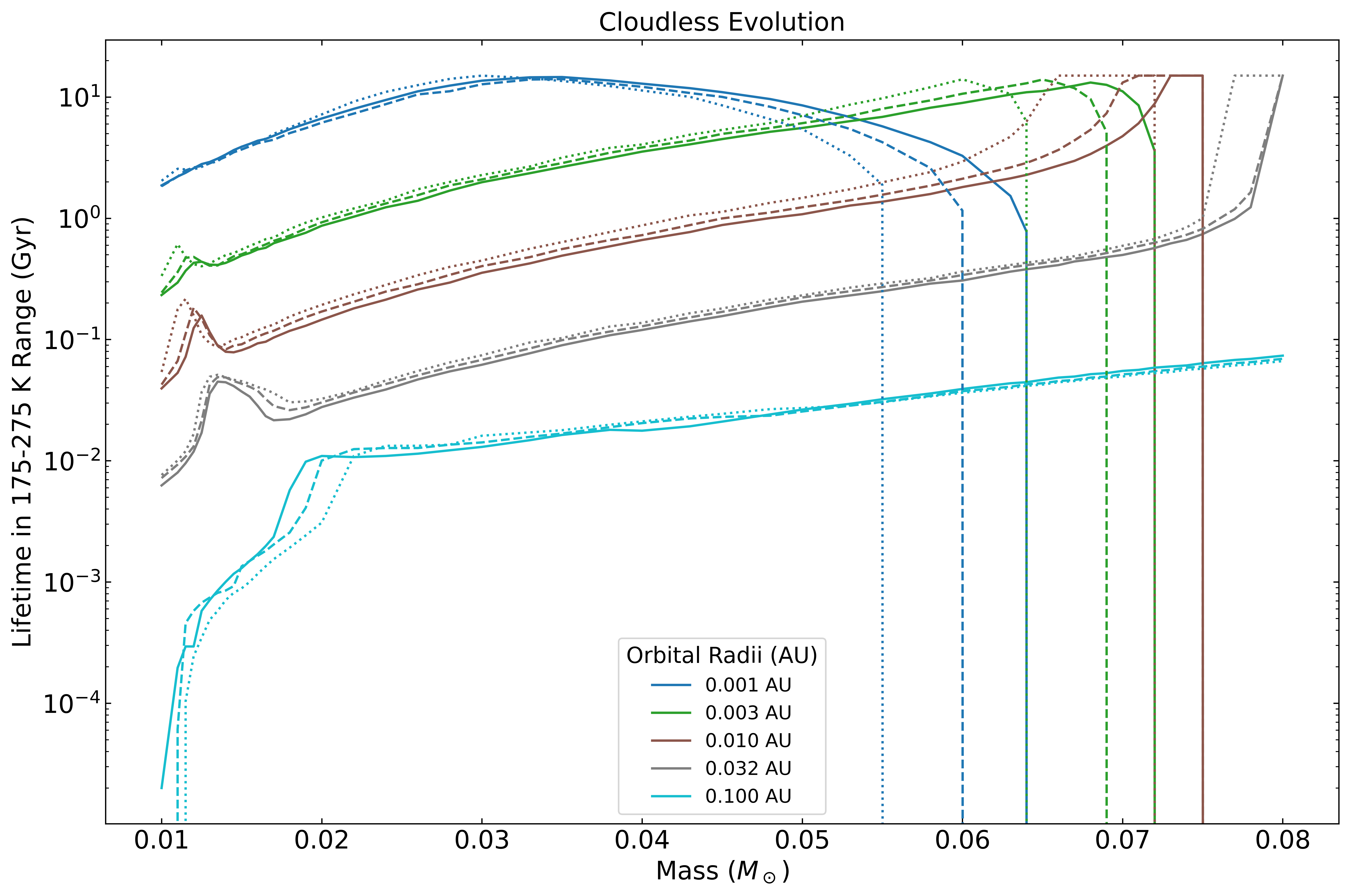}
    \caption{Duration that planets orbiting brown dwarfs remain in the HZ as a function of brown dwarf mass for the cloudless evolution. Each line represents a specific orbital radius, calculated using Equation~\ref{eq:2}. The solid line is a brown dwarf [M/H] = $-0.5$, the dashed line is a brown dwarf [M/H] = $+0.0$, and the dotted line is a brown dwarf [M/H] = $+0.5$.}
    \label{fig:lifetime_cloudless_fortney}
\end{figure*}

The slower long-term cooling of metal-rich brown dwarfs directly translates into extended HZ lifetimes. This trend is shown in Figures~\ref{fig:lifetime_cloudy_fortney}--\ref{fig:lifetime_cloudless_fortney}. Figure~\ref{fig:lifetime_cloudy_fortney} demonstrates that the duration of habitable conditions increases systematically with metallicity across most brown dwarf masses and orbital radii. Similarly, Figure~\ref{fig:cloudless_planet_temp} presents the same analysis for the cloudless evolution case, highlighting the reduced habitable lifetimes in the absence of cloud opacity. Figure~\ref{fig:lifetime_cloudless_fortney} further illustrates this trend, reinforcing the systematic increase in habitable duration with metallicity. As stated previously, clouds increase the temperature of the brown dwarfs which leads to longer planetary habitable lifetimes. A planet located at 0.01 au around a 0.0125 {M}$_\mathrm{\odot}$ brown dwarf with [M/H] = $–0.5$ remains within the HZ for approximately 202 Myr. Similarly, a planet orbiting a 0.020 {M}$_\mathrm{\odot}$ brown dwarf at the same metallicity and distance remains habitable for about 219 Myr. In contrast, at [M/H] = $+0.5$, a planet at 0.01 au around a 0.0125 {M}$_\mathrm{\odot}$ brown dwarf stays in the HZ for 273 Myr (illustrated as the dotted line in Figure~\ref{fig:lifetime_cloudy_fortney}), while a similar planet orbiting a 0.020 {M}$_\mathrm{\odot}$ brown dwarf benefits from a significantly longer habitable duration of 333 Myr.

In Figure~\ref{fig:lifetime_cloudless_fortney}, we see that metallicity significantly influences HZ lifetimes in the cloudless evolution models. At 0.01 au, a planet orbiting a 0.012 M${\odot}$ brown dwarf remains in the HZ for approximately 124 Myr at [M/H] = $-0.5$, compared to 158 Myr at [M/H] = $+0.5$. A similar trend is evident for a 0.020 M${\odot}$ brown dwarf, where the HZ lifetime increases from 145 Myr at [M/H] = $-0.5$ to 192 Myr at [M/H] = $+0.5$.

The metallicity effect is particularly noticeable in the deuterium-burning mass regime, where slower cooling enhances the deuterium “sweet spot” behavior described in Section 3.2. Higher metallicity prolongs the luminosity plateau associated with deuterium burning, thereby extending the time during which planets remain within the HZ.

In some regions of parameter space the HZ lifetime drops to zero. This occurs when a planet never cools below 275 K within 15 Gyr. Such cases are visible in Figure~\ref{fig:lifetime_cloudy_fortney} as curves that terminate abruptly at higher brown dwarf masses or smaller orbital radii. Cloudless evolution models (Figure~\ref{fig:lifetime_cloudless_fortney}) exhibit the same qualitative metallicity dependence, though the absolute lifetimes are generally longer because clouds further slow the cooling of the host.

\subsection{Inner and Outer HZ Limits}
\label{sec:3.4}
\begin{figure*}[!htb]  
    \centering
    \includegraphics[scale=0.6]{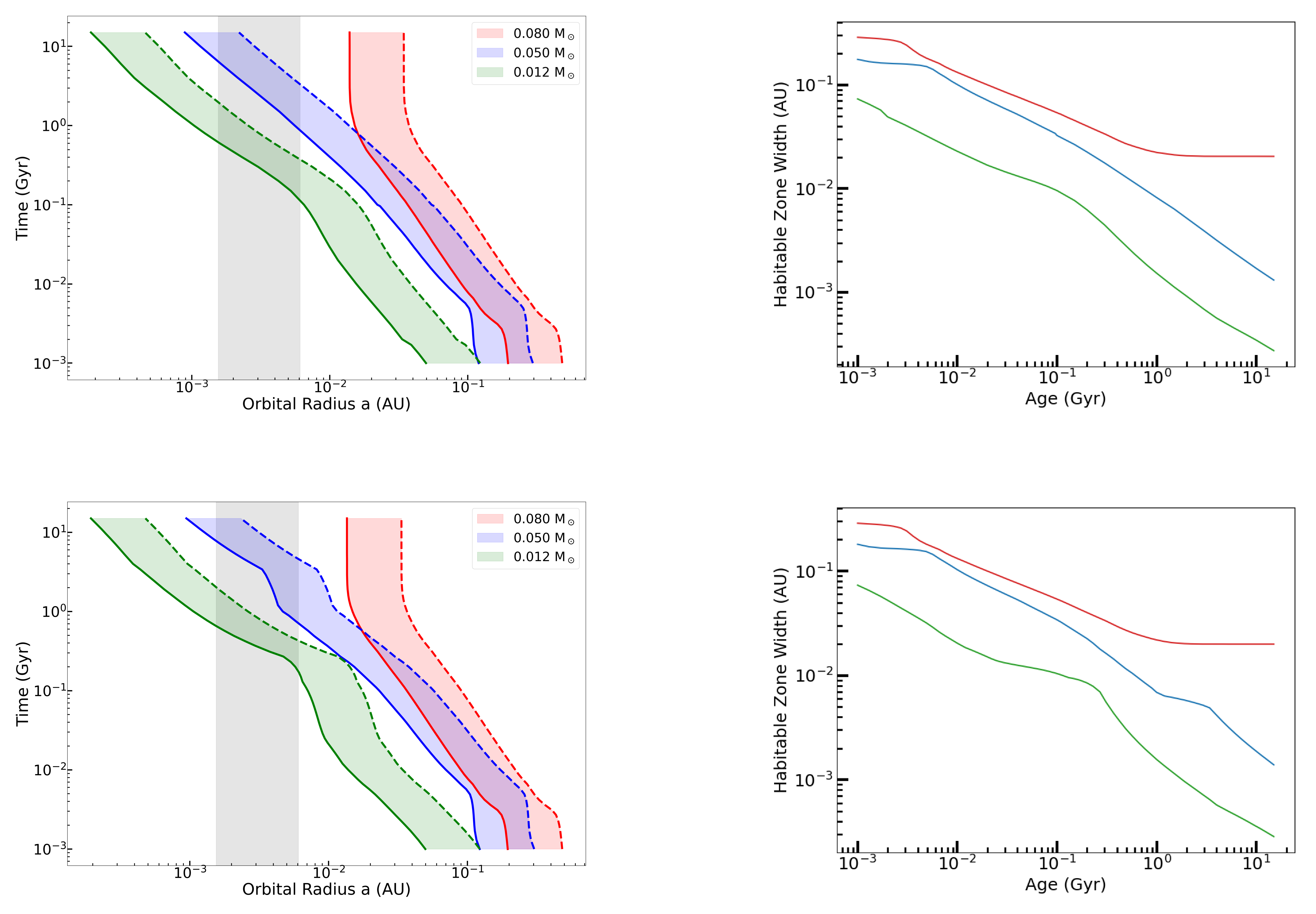}
    \caption{The evolution of the HZ for 0.012 {M}$_\mathrm{\odot}$, 0.050  {M}$_\mathrm{\odot}$, and 0.080 {M}$_\mathrm{\odot}$. The top left panel is the cloudless model, showing the boundaries for the HZ for each mass. The top right panel is the width of the HZ for the three cloudless evolution models. The bottom left and right panels are the same but are the cloudy evolution models. The solid lines are the inner HZ and the dashed lines are the outer HZ. The grey shaded region in those panels are potential corotation radii for 5-15 hour orbital periods. All of the brown dwarfs have [M/H] = $+0.0$.}
    \label{fig:inner_outer}
\end{figure*}

We visualize the trajectory of the inner and outer edges of brown dwarf HZs (both cloudless and cloudy evolution) through time in Figure~\ref{fig:inner_outer}. The more massive the brown dwarf, the further-out the HZ is at 0.001 Gyr compared to less massive brown dwarfs. It also shows that the HZ shrinks as the brown dwarf ages. For example, for cloudless 0.012, 0.050, and 0.080 {M}$_\mathrm{\odot}$ brown dwarfs, the HZ starts off with a width of 7.06$\times 10^{-2}$ au, 1.70$\times 10^{-1}$ au, and 2.77$\times 10^{-1}$ AU, respectively, at 0.001 Gyr. At 0.1 Gyr, the width of the HZ for those same brown dwarfs are 2.64$\times 10^{-4}$ au, 3.13$\times 10^{-2}$ au, and 5.20$\times 10^{-2}$ au, respectively. The figure demonstrates that the HZ location is highly time-dependent, shifting in radius substantially over the brown dwarf's lifetime. This change in width is also prominent in the cloudy evolution case, particularly for 0.012 and 0.050 {M}$_\mathrm{\odot}$.

The corotation radius is the orbital distance where a planet’s orbital period matches the rotation period of its host. Planets inside this radius orbit faster than the host’s rotation, leading to tidal torques that drive inward migration (\citealt{Bolmont_2011}). Conversely, planets outside the corotation radius orbit more slowly and experience outward tidal migration. Figure~\ref{fig:inner_outer} (gray shaded region) shows the corotation radii for brown dwarfs of 0.012, 0.050, and 0.080 M$_\odot$, computed from rotation periods ranging from 5– to 15 hr. For the M dwarf (0.080 M$_\odot$), the HZ remains entirely beyond the corotation radius at all ages.

However, for the 0.012 and 0.050 M$_\odot$ brown dwarfs, the HZ migrates inward over time and intersects the corotation radius from $\sim0.1-10$ Gyr (Figure~\ref{fig:inner_outer}). Any planet residing within what will become the HZ are expected to have already experienced strong tidal torques that drive inward orbital decay. As a result, planets may be removed through in spiral or disruption before the HZ completes its inward evolution, potentially leaving no planets remaining within the HZ at later times. Future studies must integrate both brown dwarf evolution and tidal migration with the modeling framework utilized in this work. This raises the question of which planets could actually have a lifetime in the HZ while residing within the corotation radius. Table~\ref{tab:corotation} shows the calculated corotation radii for a select few brown dwarfs to provide reference. The corotation radius was calculated using Equation~\ref{eq:3}: 

\begin{equation}
r = \left( 
\frac{\text{G} \cdot M_{\text g} \cdot P^{\,2}}
{4\pi^{2}} 
\right)^{1/3},
\label{eq:3}
\end{equation}

\noindent where G is the gravitational constant and $P$ is the rotation period.



\begin{table}[!htb]
\centering
\caption{Corotation radii for brown dwarfs with rotation periods of 5, 10, and 15 hr.}
\label{tab:corotation}
\begin{tabular}{cccc}
\hline
\noalign{\vskip 4pt}
\shortstack{Mass \\ (M$_\odot$)} &
\shortstack{$r_{\rm corot, 5hr}$ \\ (au)} &
\shortstack{$r_{\rm corot, 10hr}$ \\ (au)} &
\shortstack{$r_{\rm corot, 15hr}$ \\ (au)} \\
\hline
0.012 & 0.0016 & 0.0025 & 0.0032 \\
0.020 & 0.0018 & 0.0029 & 0.0038 \\
0.040 & 0.0023 & 0.0037 & 0.0048 \\
0.050 & 0.0025 & 0.0040 & 0.0052 \\
0.060 & 0.0027 & 0.0042 & 0.0055 \\
0.078 & 0.0029 & 0.0046 & 0.0060 \\
\hline
\end{tabular}
\end{table}


\section{Discussion}
\label{sec:4}
As shown in section~\ref{sec:3.1}, previous studies such as \citet{Lingam_2020} relied on analytic approximations rather than detailed brown dwarf evolution models, which did not accurately illustrate substellar HZ characteristics. Figure~\ref{fig:loeb} illustrates this difference: at early ages, brown dwarfs are considerably cooler than the analytic models predict and thus much less likely to lose their entire water inventory than in the analytic case, although atmospheric escape timescales from brown dwarf planets requires further study. For the first time, equilibrium tracks for planets orbiting brown dwarfs are computed using state-of-the-art models that incorporate cloud physics and provide radius and luminosity evolution from 1 Myr onward. This permits us to explore the impact of clouds and deuterium burning on brown dwarf HZs. Since clouds slow the evolution of brown dwarfs, the cloudy evolution tracks result in  orbiting planets that cool more slowly over time. 

Moreover, the transition from cloudy to cloudless states in the brown dwarf further delays planetary cooling at early ages, allowing well placed planets to maintain equilibrium temperatures in the HZ range for longer than predicted for cloudless evolution models. \citet{Lingam_2020} concluded that planets orbiting brown dwarfs with masses less than 0.030 {M}$_\mathrm{\odot}$ would not maintain habitable temperatures over significant geologic timescales. This work illustrates that if cloud formation/dissipation, deuterium burning, and realistic atmospheric opacities are incorporated into atmospheric models, planets could remain in the HZ for hundreds of millions to billions of years (depending on the brown dwarf physical parameters, orbital radius, etc.), including planets orbiting brown dwarfs less than 0.030 {M}$_\mathrm{\odot}$ (shown in Figures ~\ref{fig:lifetime_cloudless_fortney} and ~\ref{fig:lifetime_cloudy_fortney}).

Sections ~\ref{sec:3.2} and ~\ref{sec:3.3} examined the duration that planets can remain within the HZ as a function of brown dwarf properties and orbital distance. Since higher-metallicity brown dwarfs cool more slowly and remain warmer at older ages than their lower-metallicity counterparts \citep[e.g.,][]{Marley_2021}, planets orbiting metal-rich brown dwarfs spend longer periods within the HZ (see Figures~\ref{fig:cloudy_planet_temp} and ~\ref{fig:cloudless_planet_temp}). Consequently, planets orbiting metal-rich brown dwarfs spend longer periods within the HZ. This trend is evident in Figures ~\ref{fig:lifetime_cloudless_fortney} and ~\ref{fig:lifetime_cloudy_fortney}, where systems with [M/H] = +0.5 exhibit extended habitable lifetimes compared to those with lower metallicities. The HZ lifetime also depends strongly on orbital distance and host mass: planets on wider orbits generally maintain habitability longer, particularly around lower-mass brown dwarfs. Figure~\ref{fig:lifetime_cloudy_fortney} also shows the impact clouds have HZ lifetime. In the 0.001 au case shown in Figures ~\ref{fig:lifetime_cloudless_fortney} and ~\ref{fig:lifetime_cloudy_fortney}, planets with masses above approximately 0.050 M$_\odot$ exhibit longer lifetimes in Figure~\ref{fig:lifetime_cloudless_fortney} than in Figure~\ref{fig:lifetime_cloudy_fortney}. This difference arises because the brown dwarfs at these masses are hotter (due to the formation of clouds), resulting in planetary temperatures exceeding 275 K.

Section ~\ref{sec:3.4} examines the evolution of the inner and outer HZ boundaries for both cloudless and cloudy brown dwarf models. In all cases, the HZ contracts and migrates inward as the brown dwarf cools over time. The inward migration and contraction of brown dwarf HZs shown in Figure~\ref{fig:inner_outer} have important consequences when coupled with tidal evolution. For the 0.012 and 0.050 M$\odot$ brown dwarfs, the HZ intersects the corotation radius at late times, implying that planets residing in the HZ may experience strong inward tidal migration and be removed before long-term habitability can be sustained. In contrast, for the 0.080 M$\odot$ M dwarf, the HZ remains exterior to the corotation radius throughout its evolution.

\begin{figure*}[!htb]  
    \centering
    \includegraphics[scale=0.3]{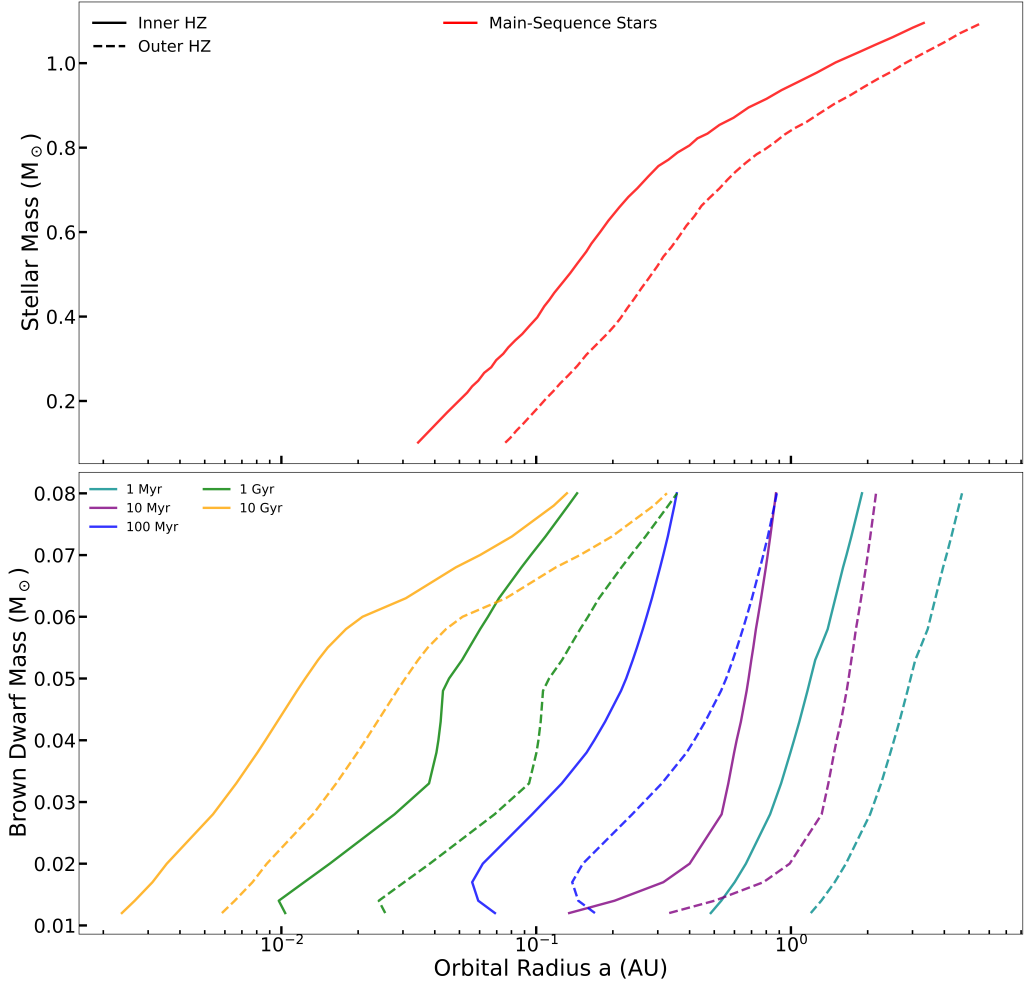}
    \caption{The top panel are the inner and outer HZ limits for main-sequence stars as a function of stellar mass, according to \citep[][]{Kopparapu_2013}. The bottom panel is the same but for cloudy evolution brown dwarfs. The solid lines are the inner HZ and the dashed lines are the outer HZ. The difference between objects at the hydrogen-burning threshold arises because \citet{Kopparapu_2013} employed a full radiative--convective atmospheric model, whereas this study assumes a constant planetary Bond albedo of 0.25. All brown dwarfs have [M/H] = +0.0.}
    \label{fig:kopparapu}
\end{figure*}

Similarly, Figure~\ref{fig:kopparapu} presents a comparison of the HZs of cloudy brown dwarfs ([M/H] = 0.0) and main-sequence stars from \citep[]{Kopparapu_2013}. The HZ boundaries for brown dwarf ages 1 Myr, 10 Myr, 100 Myr, 1 Gyr, and 10 Gyr are shown. A key distinction evident in Figure~\ref{fig:kopparapu} is that the HZ width for the cloudy evolution brown dwarfs varies only weakly with mass at younger ages, whereas it decreases substantially along the main sequence. This difference reflects the contrasting luminosity–mass dependencies of the two regimes: main-sequence luminosities scale steeply with mass and stellar effective temperature, causing both the inner and outer HZ limits to shift outward rapidly with increasing stellar mass and to converge at the low-mass end. 

It is also important to note that connecting the hydrogen-burning limit of main-sequence stars to the brown dwarfs shifts the brown dwarf HZ farther outward, meaning the corresponding temperatures are higher. This occurs because \cite[][]{Kopparapu_2013} adopt lower albedos (and used a full radiative-convective model), which trigger a runaway greenhouse at lower levels of insolation and therefore place the inner edge of the HZ at smaller orbital distances. The planetary albedo vs. surface temperature of planets orbiting a variety of main-sequence stars is shown in Figure 6 in \cite[][]{Kopparapu_2013}.

In contrast, the shallower luminosity–mass relation of brown dwarfs produces nearly parallel HZ boundaries, yielding a relatively constant absolute width at younger ages. The systematically larger orbital distances of the main-sequence HZs are a direct consequence of their higher effective temperatures and bolometric luminosities, which shift the equilibrium temperature range for liquid water to wider separations. It is also clear that the HZ boundaries shift inward with time since brown dwarfs cool with age. Overall, brown dwarf HZs sweep through a large range of radii and Figure~\ref{fig:kopparapu} illustrates how complex the evolution is.


Although no rocky exoplanets have yet been discovered around brown dwarfs, studies have placed theoretical constraints on the occurrence rates of short-period planets in such systems (e.g., \cite{2017MNRAS.464.2687H}. The transit method is particularly favorable for detecting rocky planets orbiting brown dwarfs: an Earth-sized planet transiting a brown dwarf with a radius of \(0.1\,R_{\odot}\) produces a transit depth of \(\sim8400\) ppm, compared to only \(\sim84\) ppm for the same planet orbiting a G-type star. This large contrast underscores the strong potential for future searches targeting HZ planets around brown dwarfs. In future work, we will examine how the unique spectral energy distributions of brown dwarfs shape planetary climate by applying a 1D radiative–convective model to determine the inner (runaway greenhouse) and outer HZ limits for Earth-like planets \citep[][]{windsor2022radiativeconvectivemodelterrestrialplanets}.










However, the variability of the brown dwarf can cause the transit data to be extremely noisy. This is also seen in M-dwarf variability. There have been many studies analyzing the high activity and its effect on transits in M dwarfs \citep[e.g.,][]{Yaptangco_2025, miyakawa2022colordependencetransitdetectability}. Similar studies to those listed could be a point of research for brown dwarfs as well. Considering that brown dwarfs have opposite evolution as main-sequence stars could provide another interesting window to the possibilities of habitability beyond Earth. 

\cite{Barnes_2013} highlighted fundamental challenges to habitability around cooling brown dwarfs (including tides). Our work further demonstrates that when cloud evolution and metallicity-dependent cooling are included, HZ lifetimes can be substantially longer than previously assumed. In contrast to  \cite{Bolmont_2018}, who emphasized that inward HZ migration coupled with tidal torques near the corotation radius may rapidly eliminate potentially habitable planets around brown dwarfs, our results demonstrate that when realistic brown dwarf evolutionary physics is taken into account, planets located beyond $\sim 0.0016-0.006$ au, depending on the mass and corotation radius of the host brown dwarf, can both remain outside the corotation radius and persist within the HZ for extended durations (shown in Table~\ref{tab:corotation}).


\section{Conclusions}
\label{sec:5}
In this work, we calculated new revised equilibrium temperatures for planets orbiting complex and variable brown dwarfs.
\begin{itemize}
    \item We find that brown dwarfs are significantly less luminous at young ages than predicted by analytic approximations used in some recent brown dwarf HZ studies \citep[][]{Lingam_2020}. Those approximations neglect deuterium burning and cloud formation and dissipation, both of which play critical roles in determining accurate HZ boundaries and their evolution.
    \item Deuterium burning and cloud formation and dissipation substantially extend HZ lifetimes for planets at specific orbital distances by maintaining those planets within the HZ while these processes operate.
    \item Brown dwarfs with higher metallicities support longer HZ lifetimes because increased atmospheric opacity slows their cooling.
    \item As brown dwarfs age, their HZs contract by orders of magnitude, depending on the brown dwarf mass.
\end{itemize}

This work provides a framework to guide future searches for habitable planets orbiting brown dwarfs, expanding the search for life beyond main-sequence stars.

\begin{acknowledgments}
K.J.S. would like to thank her funding sources of this work: the University Fellows Program at the University of Arizona and the Carson Fellowship through the Lunar and Planetary Laboratory. K.J.S. and M.M. would also like to thank the reviewer for the insightful comments that improved this manuscript. 
\end{acknowledgments}

\begin{contribution}
All authors contributed equally to this work.
\end{contribution}

%
\facilities{None}

\section{Software and third party data repository citations} 
\noindent Cloudless models ({\cite{Marley_2021}), Cloudy models (\cite{morley2024sonorasubstellaratmospheremodels}), and brown dwarf models at early ages (\cite{2025arXiv251008694D}).

\bibliography{paper}{}
\bibliographystyle{aasjournalv7}



\end{document}